\documentclass[authoryear]{elsarticle}

\journal{XXIII Congresso Brasileiro de Autom\'atica - XXIII Brazilian Congress of Automatica}

\usepackage[hyperindex,pdftex]{hyperref}
\usepackage{graphicx}
\usepackage{amssymb,amsmath}
\usepackage{enumitem}
\usepackage{mathtools}
\usepackage{multirow}
\usepackage{xcolor}

\newdefinition{remark}{Remark}

\begin{document}

\begin{frontmatter}

%\usepackage{hyperref}

%\newcolumntype{P}[1]{>{\centering\arraybackslash}p{#1}}
%\newcolumntype{M}[1]{>{\centering\arraybackslash}m{#1}}

%opening
\title{Optimal Control Concerns Regarding the COVID-19 (SARS-CoV-2) Pandemic in Bahia and Santa Catarina, Brazil}

\author[adufsc]{Marcelo M. Morato}
\ead{marcelomnzm@gmail.com}
\author[adUAL]{Igor M. L. Pataro}
\author[adUFBA]{Marcus V. Americano da Costa}
\author[adUFSC]{Julio E. Normey-Rico}

\address[adUFSC]{Renewable Energy Research Group (\emph{GPER}), Department of Automation and Systems (\emph{DAS}),Federal University of Santa Catarina (UFSC), Florian\'opolis, Brazil.}
\address[adUFBA]{Department of Chemical Engineering (\emph{DEQ}), Federal University of Bahia, 02 Professor Aristides Novis St., Salvador, BA-40210910, Brazil.}
\address[adUAL]{CIESOL, Department of Informatics, University of Almería, Ctra. Sacramento s/n 04120, Almería, Spain}

\begin{abstract}
The COVID-19 pandemic is the profoundest health crisis of the $21^{\text{rst}}$ century. The SARS-CoV-2 virus arrived in Brazil around March, 2020 and its social and economical backlashes are catastrophic. In this paper, it is investigated how Model Predictive Control (MPC) could be used to plan appropriate social distancing policies to mitigate the pandemic effects in Bahia and Santa Catarina, two states of different regions, culture, and population demography in Brazil. In addition, the parameters of Susceptible-Infected-Recovered-Deceased (SIRD) models for these two states are identified using an optimization procedure. The control input to the process is a social isolation guideline passed to the population. Two MPC strategies are designed: a) a centralized MPC, which coordinates a single control policy for both states; and b) a decentralized strategy, for which one optimization is solved for each state. Simulation results are shown to illustrate and compare both control strategies. The framework serves as guidelines to deals with such pandemic phenomena.
\end{abstract}


\end{frontmatter}
%\linenumbers

%\newpage

\section{Introduction}
\label{sec1}

 The COVID-19 pandemic is the global health crisis of the $21^{\text{rst}}$ century. The SARS-CoV-2 virus was first registered in humans in Wuhan, China by December 2019. The virus causes a severe acute respiratory syndrome which can become potentially fatal. The contagion spreads rapidly and efficiently; by mid-Jun, this pandemic had already spread to nearly all countries of the world, killing more than $410000$ people. To address and mitigate the pandemic, global scientific efforts are necessary \citep{bedford19}; since vaccines are previewed to be ready only by mid-$2021$, COVID-19 has posed an unique question and the majority of countries have adopted social distancing measures, seeking to avoid its spread \citep{Adam2020}.

Much more than presenting drastic effects on health systems, social and economical backlashes are already felt by many countries; this is especially evident in countries with larger social inequalities, such as Brazil, which is now the second leading country in numbers of cases and deaths. The effects of the virus on populations with poorer access to health systems and sanitation facilities\footnote{A very illustrative example of these differences can be seen in the city of S\~ao Paulo: the city hall released a technical note by the end of April stating that the observed mortality rate is $10$ times larger in neighborhoods of the city with worse social conditions and precarious housing. See \url{https://www.prefeitura.sp.gov.br/cidade/secretarias/upload/saude/PMSP_SMS_COVID19_Boletim\%20Quinzenal_20200430.pdf}.} are strikingly harder  \citep{san2020covid}. 

In this paper, the Brazilian context is taken into account \citep{werneck2020covid}: Brazil is a continent-sized country with $26$ federated states, which have been choosing different social distancing measures since mid-March. The federal government is reluctant to implement nation-wide policies, claiming that the negative economic effects are too steep and that social distancing is an erroneous choice \citep{THELANCET20201461}. Thus, the government suggests that the economy cannot stop and that herd immunity could be a solution to this pandemic. However, the expectations present on the recent literature suggest catastrophic scenarios for the next months \citep{rocha2020expected, morato2020optimal}. To account for locations which have been following very different paths regarding COVID-19, the data from two states are considered: (\textit{i}) Bahia (BA), which lies on the northeast sea-side and is larger than Spain (in total surface), and (\textit{ii}) Santa Catarina (SC), which is in the south of the country and is three times larger than Belgium. Also, it is important to highlight that it is expected different social behaviour in these states due to their history, culture, and other aspects.

\begin{figure*}[]
	\centering
		\includegraphics[width=\linewidth]{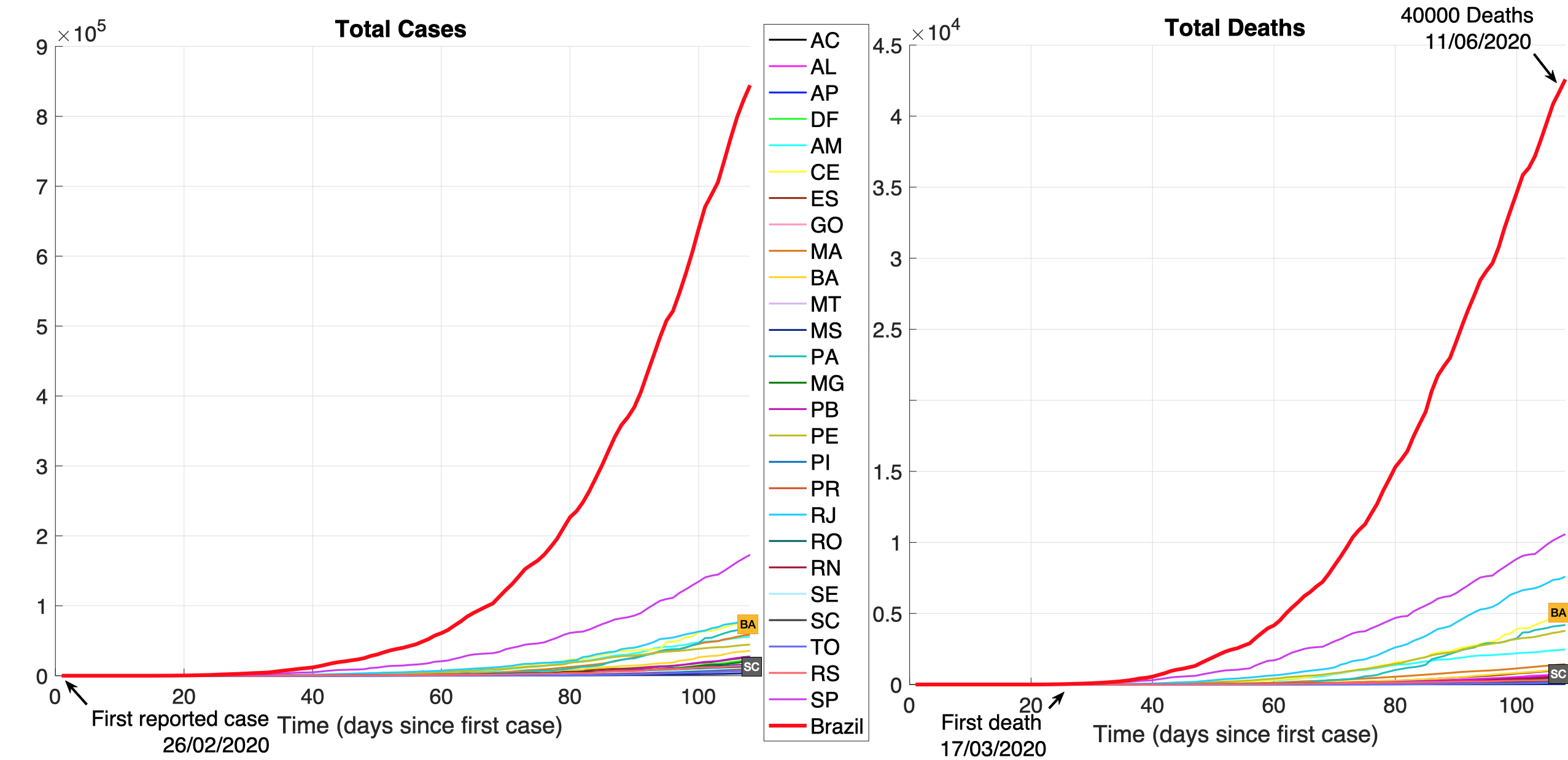}
                \caption{COVID-19 in Brazil.}
	\label{perStatePast}
\end{figure*}

To better illustrate the Brazilian scenario, in Figure \ref{perStatePast}, the available data regarding the total number of SARS-CoV-2 confirmed cases and deaths is shown. Note that as of $22/06/2020$, Brazil counts over $50000$ deaths due the SARS-CoV-2 virus. 

%\FloatBarrier

It should be mentioned that the idea behind social distance is to prevent health systems from becoming saturated due to large amounts of COVID-19 patients being treated at the same time. Therefore, with social distancing policies, the health systems do not have to deal with hospital bed shortages associated with a large peak of infections, since the demands for treatment become distributed over time. Even though a strong public health system is available in Brazil, as of April $30$, many states were already exhibiting a near-collapse situation, with over $95\, \%$ of Intense Care Unit (ICU) hospital beds occupied with COVID-19 patients. Since BA and SC are very different states in demographic terms, while both applied strict social isolation at first (around March), today they face different situations: BA has roughly $70\, \%$ of ICU occupancy, whereas of SC has a lower rate, estimated at $22 \, \%$. 

A fundamental issue regarding social distancing is to perform such intervention at the correct time and for the correct duration. Well-designed social distance policies should help mitigating the contagion and thus avoiding the saturation of the heath systems and to minimizing their social and economic side-effects. 

Motivated by the previous discussion, the problem of how optimal control can be used to formulate adequate social distancing policies, regarding BA and SC, is investigated in this work. The Model Predictive Control (MPC) \citep{camacho2013model} framework is used, since it can conveniently consider the effect of lockdown/quarantine measures as the constraints 
of a minimization problem (regarding the number of infection). Furthermore, the differences regarding the use of a centralized MPC scheme that generates the same control law for both states, and the use of distributed MPC approach are compared, solving two separate procedures with individual laws for each state.

Based on on a Susceptible-Infected-Recovered-Deceased (SIRD) model adjusted for the COVID-19 pandemic from \citep{bastos2020modeling}, which embeds the effects of social distancing measures (Section \ref{sec2}), the main contributions of this paper are the following:
\begin{itemize}
    \item An optimization procedure is developed in order to minimize a Least-Squares criterion and estimate the parameters of the virus infection/spread model, considering both states (BA, SC). Uncertainty in the available datasets is considered in the identification (Section \ref{sec3});
    \item Based on the obtained models, two different MPC strategies are designed in order to determine when to apply (or not) social distancing policies (Section \ref{sec4});
    \item Simulation results illustrate the results obtained with both strategies; discussions are drawn in order to evaluate how optimal control can be used to guide social distancing in pandemic situations (Sections \ref{sec5} and \ref{sec6}). The discussion is formally based on comparisons regarding the COVID-19 spread in both states of BA and SC and how the different control strategies can address the goal of mitigating the viral spread.
\end{itemize}

\section{SARS-CoV-2 Pandemic Spread Model}
\label{sec2}

Recent literature \citep{peng2020epidemic,kucharski2020early} discusses how the the infection rate and evolution dynamics of the SARS-CoV-2 virus can be adequately described by Susceptible-Infected-Recovered-Deceased (SIRD) models.
In this Section, the SIRD model from \citep{keeling2011} is detailed. This model is adapted with a dynamic variable which models the population's response to isolation policies, as proposed in \citep{bastos2020modeling,morato2020optimal}.

\subsection{SIRD Epidemiological model}

The SIRD equations describe a contagion spread in a population which is split into four non-intersecting classes:
\begin{itemize}
    \item Susceptible people\footnote{In this paper, we do not consider the effects of demographic variations. Despite recent discussion regarding the possibilities of reinfection \citep{del2020covid}, we assume that the recovered individuals will not be reinfected (at least for simplicity purposes), i.e. an individual does not contract the disease twice.} $S(t)$, who are prone to contract the virus;
    \item Infected individuals $I(t)$, which are currently sick;
    \item Recovered people $R(t)$, who have already recovered from the SARS-CoV-2;
    \item and Deceased individuals $D(t)$, who have died due to the contagion. 
\end{itemize}   

 The SIRD model follows:
\begin{equation}
    \begin{array}{rcl}
    \displaystyle \frac{dS(t)}{dt} & = & \displaystyle - (1-\psi(t)) \frac{\beta I(t) S(t)}{N(t)} \text{,}\\[3mm]
    \displaystyle \frac{dI(t)}{dt} & = & \displaystyle (1-\psi(t)) \frac{\beta I(t) S(t)}{N(t)} - \frac{\gamma I(t)}{1 - \rho} \text{,} \\[3mm]
    \displaystyle \frac{dR(t)}{dt} & = & \displaystyle \gamma I(t) \text{,} \\[3mm]
    \displaystyle \frac{dD(t)}{dt} & = & \displaystyle \frac{\rho}{1-\rho} \gamma I(t) \text{,}\\[5mm]
    \end{array}\;\;\;\textrm{\bf [SIRD]}
\label{eq:SIRDead}
\end{equation}

wherein the parameter \(\beta\) stands for the probality of disease transmission per contact times the number of contacts per unit time \(t\); \(\gamma\) stands for the recovery rate, determining the amount of individuals that ``leave'' from the infected class, and the parameter $\rho$ denotes the observed mortality rate. Social distancing measures expressed through $\psi (t)$, denoting the average amount of people circulating freely, i.e. $\psi \,=\, 1$ stands for a complete isolation condition ($100 \, \%$ quarantine, the amount of contacts are reduced to zero), whereas $\psi \,=\, 0$ means no social distancing.

The size of the total population exposed is denoted $N(t)$; it holds that $N(t) = N_0 - D(t)$, in which $N_0$ is the initial population size (prior to the contagion). However, in this work, it is assumed a constant $N(t)$, that is, natural deaths balance the newborns. Moreover, the term \(\beta I(t)/N(t)\) stands the average number of contacts sufficient for viral transmission to one susceptible individual, per unit of time; and \((\beta I(t)/N(t))S(t)\) gives the total the number of new cases with respect to the amount of susceptible individuals (they are ``available for infection''), per unit of time. 

Another essential information in epidemiology theory is the basic reproduction number, usually denoted by $R_0$. This factor is able to measure the average  potential transmissibility of the disease. In practical analysis, it represents how many expected cases could be generated by a single primary case in a population in which all invidious are susceptible. In point of view of dynamic systems, $R_0$ represents the epidemic velocity. If $R_0> 1$ the infection is spreading and the number of infected people increases, which happens at the beginning of the epidemic, otherwise, if $R_0<1$ it means more individuals ``leave'' from the infected class, either recovering or dying due to fatal cases, and thereby the epidemic ceases. The reproduction number $R_0$ is affected by different factors, including biological characteristics from the virus itself, and governments policies to control the number of susceptible people, which can be reduced by social distancing. 

To calculate $R_0$, it is assumed that at the beginning of the pandemic,  $S\approx N$. Considering the parameters $\beta$, $\gamma$ and $\rho$ from Equation \ref{eq:SIRDead} related to the rate of infected invidious, $R_0$ can be approximated as follow:
\begin{eqnarray} %\label{eq:err}
\nonumber R_0 \approx \frac{\beta(1-\rho)}{\gamma}
\end{eqnarray}

\subsection{Social Distancing Variable}

To take into account the effect of public health policies, enacted by local governments to mitigate the effects of the COVID-19 pandemic, we follow the method proposed in \citep{bastos2020modeling,morato2020optimal}: including a dynamic model for $\psi (t)$ to the SIRD dynamics. We note that that $\psi$ models not only social isolation, but also incentives to use of masks and other measures which contain the contagion spread. The dynamics are:
\begin{eqnarray} \label{eq:Psi}
\frac{d\psi(t)}{dt} &=&  \frac{1}{\varrho} (u(t)\psi_{\infty} - \psi(t)) \text{,}
\end{eqnarray}
in which $u(t)$ is the control variable defined within $[0\, , \, 1]$ that sets social distancing measures. Note that $\psi(t)$ converges to $\psi_\infty$ with a settling time of $3\varrho$ for $u = 1$. It follows that $\psi_\infty$ is a factor which stands for the  maximal observed effect of pandemic policies. We note that for larger values of $\psi_\infty$ (closer to $1$), when hard quarantine measures are enacted ($u$ closer to $1$), the SIRD model dynamics (with $\psi(t) \to \psi_\infty$) are slowed down, exhibiting a smaller peak of infections and number of deaths. 

\section{Identification Procedure}
\label{sec3}

For the sake of practical purposes, model parameters and distancing policies are estimated using real data employed in a similar technique to the estimation scheme presented by \citep{bastos2020modeling}. 

Ordinary Least Square (OLS) method is applied for estimating parameters $\beta$, $\gamma$ and $\rho$ from Equation \ref{eq:SIRDead}. For the algorithm developed here, it is used official data provided by Ministry of Health of Brazil from the first case registered in each studied Brazil state: March 13, 2020 to Santa Catarina, and March 06, 2020 to Bahia, until the last available data on June 16, 2020.

First, it is considered the SIRD model with no public health measure or policy adopted by the government, leading, thus, $\psi(t)=0$. Furthermore, for the sake of simplify formulation of the optimization algorithm, the differential equations $dI(t)/ dt$ and $dD(t)/ dt$ are modified in order to find a linear formulation with respect to parameters, which leads to the following expressions:
\begin{eqnarray} 
       \frac{dI(t)}{dt} &=& \frac{\beta I(t) S(t)}{N(t)} - \gamma I(t) -\alpha I(t) \text{,} \\
       \frac{dD(t)}{dt} &=& \alpha I(t) \text{,}
\end{eqnarray}
 \noindent for $\alpha = \rho  \gamma/(1-\rho)$. 

The optimization problem is formulated taking into account the minimum square error between real data (as disclosed by the Brazilian Ministry of Health) and the estimated SIRD dynamic model in function of $\beta$, $\gamma$ and, now, $\alpha$. For each variable I(t), R(t) and D(t), the error is calculated as follows:
\begin{eqnarray} %\label{eq:err}
\nonumber Er_{I} &=& (I(t) - \hat{I}(t,\beta,\gamma,\alpha))^2\text{,} \\ \nonumber
Er_{R} &=& (R(t) - \hat{R}(t,\beta,\gamma,\alpha))^2 \text{,}\\ \nonumber
Er_{D} &=& (D(t) - \hat{D}(t,\beta,\gamma,\alpha))^2 \text{,}
\end{eqnarray}
\noindent for which the variables $\hat{I}, \hat{R}, \hat{D}$ are estimated according to the SIRD model equations. Finally, the complete optimization problem is formulated as follows:

\begin{eqnarray}
\nonumber
    \underset{\beta,\gamma,\alpha}{\min}  J &&= \underset{\beta,\gamma,\alpha}{\min}  \sum_{i=ti}^{i=t_i+t_{opt}} \left(k_1Er_I(i) + k_2 Er_R(i) + k_3 Er_D(i)\right) \text{,} \\ \label{eq:opt}
    \text{s.t.:} && 0 \leq \beta \leq 0.65 \text{,}\\ \nonumber
    && 0 \leq \gamma \leq 0.7 \text{,}\\ \nonumber
&& 0 \leq \alpha \leq 0.2 \text{.}
\end{eqnarray}

The optimization begins with initial conditions $\beta = 0.5$, $\gamma = 0.5$ and $\alpha = 0.1$. The tuning parameters $k_1$, $k_2$ and $k_3$ are taken as positive weighting values, used to normalize the magnitude order of the total cost with respect to variables $Er_I$, $Er_R$ and $Er_D$. The optimization window is considered between an initial time instant $t_i$ and a terminal instant $t_{opt}$. 

To solve this problem, we consider a fixed optimization horizon in this work, since the available data may not represent the real trend of the epidemic dynamics. This is due to the fact that uncertainties are present in the first data available related the reported cases and, hence, we assume that this information could possibly deteriorate the overall parameters estimation and the obtained model prediction. Moreover, initial data is usually embed substantial variations in the number of cases reported due to the absence of testing when the viral spread starts. Also, as a natural consequence of pandemic, infections, recovered and mortality rates start with strong variations at the beginning of the spread, until convergence to a steadier behavior is seen (which is the phenomena currently observed in Brazil).

The algorithm is performed over a horizon window ($t_{opt}$), which is smaller than the total available amount of data and, thus, determines the optimal constant parameter values for the SIRD model regarding this window. We have found that the best model-data fitting results were achieved with an optimization windows between $5$ and $10$ days, which is coherent with the viral dynamics (average incubation period is of $5$ days, at most $14$). Hence, we identify parameters$\beta$, $\gamma$ and $\alpha$ for each window $t_{opt}$. The procedure starts with $t_i=1$ as the first day of available data within the considered window. In the next optimization loop ($t_i=t_i+t_{opt}+1$), the optimal parameters identified on the last window instant ($\beta_{opt}$, $\gamma_{opt}$ and $\alpha_{opt}$) are set as initial conditions for the current window, and the loop continues until the last available data. The procedure has a moving-window strategy.

%\begin{figure}[htb]
%	\centering
%		\includegraphics[width=\linewidth]{./img/optalgo.png}
%                \caption{Optimization Algorithm}
%	\label{optalgo}
%\end{figure}
%\FloatBarrier

\section{An Optimal Social Distancing Method}
\label{sec4}

Based on the SIRD model detailed in Sec. \ref{sec2} and the parameter estimates found through the optimization procedure in Sec. \ref{sec3}, two different optimal control procedures to guide social distancing policies in Brazil are now proposed. These procedures are set within an MPC framework, in centralized and distributed paradigms, as detailed in the sequel.

It can be noted that the MPC operates in a discrete-time manner. Therefore, since new measurements of infections and deaths are available every day, the SIRD model is Euler-discretized with $T_s \, = \, 1$ day. The discrete instants are denoted as $k = t/T_s$.

\subsection{Optimization Goal}

The MPC formulation is designed through a minimization problem, where performance goals are delimited as quadratic functions. Regarding the COVID-19 situation, the control objective is evident: minimize the number of active infections ($I(k)$) while reducing the social distancing ($u(k)$). Note that countries that implement rigid social distancing measures have seen devastating economic effects and thus this measure should be kept for the smallest time possible; recent papers elaborate on this issue, e.g. \citep{NBERw26882}.

\subsection{Process Constraints}

One cannot expect to increase of decrease social isolation instantaneously. As observed in practice, the population takes some time to respond to new social isolation measures, adapting to the enacted paradigm. Therefore, in consonance with the dynamic Equation \eqref{eq:Psi} and with real isolation policies put in practice in Brazil, we consider that the control action $u$ can vary $\pm 0.05$ per day, which means that actual isolation factor will increase/decrease with a rate of, at most, $5\, \%$/day.

We note that this is a preliminary assumption, since the actual implemented social isolation policy should be a "translation" of the control signal $u$ to feasible actions. This actions could represent different guidelines, for instance: a total isolation, with no body leaving their houses (for $u \, = \, 1$), a partial isolation, with people allowed to leaved only for short periods, with masks (for $u \, = \, 0.9$), and so forth, until a total "relaxed" condition (for $u \, = \, 0$).

\subsection{Centralized MPC}

Bearing in mind the control goal and constraints above, the first MPC procedure proposed is the set as a single, centralized MPC (CMPC) algorithm which takes into account the evolution of the contagion in the whole country (considering data-sets per state) and thereby determines a single control action $u$ which guides social isolation policies for all states. Such optimization can be expressed as follows:
\begin{eqnarray}
\label{CMPC}
    \underset{U(k)}{\min}  \, J_{CMPC} &&= \underset{U(k)}{\min} \,  \sum_{j}\sum_{i=1}^{N_p} \frac{\left(I_j(k+i)^Tq_II_j(k+i)\right)}{n_j^2} \\ \nonumber && \,\,\,+\sum_{i=1}^{N_p}\left( u(k+i-1)^Tq_uu(k+i-1)\right) \text{,} \\ \nonumber
    \text{s.t.:} && \text{Discrete SIRD Models} \,\,\, \forall \, i \in \, \mathbb{N}_{[1\, , \, N_p]} \,\text{,} \\ \nonumber
    && 0 \leq u(k+i-1) \leq 1 \,\text{,}\\ \nonumber
    && -0.15 \leq u(k+i) - u(k+i-1)\leq 0.15 \,\text{,}\\ \nonumber
&& 0_{2\times 1} \leq  \begin{bmatrix}
I_j(k+i) \\ 
D_j(k+i)\\
\end{bmatrix}  \leq \begin{bmatrix} n_j^{ICU} \\ n_j \end{bmatrix} \, \forall \, j \,\text{,}
\end{eqnarray}
\noindent wherein $N_p$ is a prediction horizon, the sub-script $j$ indicates the state (i.e. $I_{BA}$ stands for the infections in Bahia), $n_j$ stands for the total population size of the $j$-th state, $n_j^{ICU}$ represents the total ICU beds available in the state and $U(k)$ represents the sequence of control actions inside the prediction horizon, i.e. $U(k) \,=\, \text{col}\{u(k) \, u(k+1) \, \dots \, u(k+N_p-1)\} $. The weights $q_I$ and $q_u$ determine the trade-off between conflicting objectives of minimizing the spread and reducing social isolation.

\subsection{Distributed MPC}

A distributed MPC (DMPC) formulation is much like the one in Eq. \eqref{CMPC}, but considers the SIRD model for just a single state, and, thereby, finds an individual control law for the referenced state. The optimization procedure is the following:
\begin{eqnarray}
\label{DMPC}
    \underset{U(k)}{\min}  \, J_{DMPC}^j &&= \underset{U(k)}{\min}\,  \sum_{i=1}^{N_p} \frac{\left(I_j(k+i)^Tq_II_j(k+i)\right)}{n_j^2} \\ \nonumber && \,\,\,+\sum_{i=1}^{N_p}\left( u(k+i-1)^Tq_uu(k+i-1)\right) \text{,} \\ \nonumber
    \text{s.t.:} && \text{Discrete SIRD Model ($j$)} \,\,\, \forall \, i \in \, \mathbb{N}_{[1\, , \, N_p]} \,\text{,} \\ \nonumber
    && 0 \leq u(k+i-1) \leq 1 \,\text{,}\\ \nonumber
    && -0.15 \leq u(k+i) - u(k+i-1)\leq 0.15 \,\text{,}\\ \nonumber
&& 0_{2\times 1} \leq  \begin{bmatrix}
I_j(k+i) \\ 
D_j(k+i)\\
\end{bmatrix}  \leq \begin{bmatrix} n_j^{ICU} \\ n_j \end{bmatrix} \, \text{.}
\end{eqnarray}

The CMPC approach, regarding Brazil, would stand for a single social isolation guide to all $26$ states, where a DMPC design would represent social isolation defined through individual state guidelines. For simplicity, although the whole approach presented here can be extrapolated for other scenarios, this work considers only the states of Bahia and Santa Catarina in the following results.

\begin{remark}
In recent papers \citep{morato2020optimal,kohler2020robust}, the issue of MPC regarding COVID-19 has also been discussed. Notice that the differences between the CMPC/DMPC formulations presented in this paper to those in the references are the following: \textit{(i)} the control signal presented by \cite{kohler2020robust} is a factor that multiplies the contagion transmission factors $\beta$ and $\gamma$, while in this paper and in work of \cite{morato2020optimal}, it goes through a dynamic model regarding $\psi$; \textit{(ii)} both previous papers consider uncertainty and approach the problem using a robust design procedure, which is out of the scope of this paper, since, as of today, there is enough data to find consistent parameters (the references date at a time when no so much data was available); and \textit{(iii)} the DMPC/DMPC approaches consider slew rate constraints on the control signal $u$, which had not yet been tested (the previous references considered that the social isolation reference could vary arbitrarily at each future sample $k+i$).
\end{remark}

\section{Main Results}
\label{sec5}

 We proceed by depicting the results concerning the identification procedure and the obtained control results. The following results were obtained with the aid of Matlab software, Yalmip toolbox and fmincon solver.

\subsection{Model Identification}
The SIRD identification procedure is performed through the optimization given in Eq. \eqref{eq:opt}, with the weights presented in Table \ref{tabelOpti}. The identification is performed considering a window of $6$ days, both for BA and SC states.

\begin{remark}
At the beginning of the pandemic in SC, there was a considerable lack of reported recovered cases (until May $5^{th}$), inconsistencies regarding the active infections and $R_0$. However, this does not affect the overall forecasts due to the moving window optimization strategy and, thereby, the control strategy methodology.
\end{remark}

\begin{table}[htbp]
    \caption{\label{tabelOpti} Optimization Weights.}
    \centering
	\begin{tabular}{|c | c c c |} 
		\hline 
        Parameter & $k_1$ & $k_2$ & $k_3$ \\ \hline
        Value & $1$ & $10$ & $2$ \\ \hline
    \end{tabular}
\end{table}

\begin{figure}[htb]
	\centering
		\includegraphics[width=\linewidth]{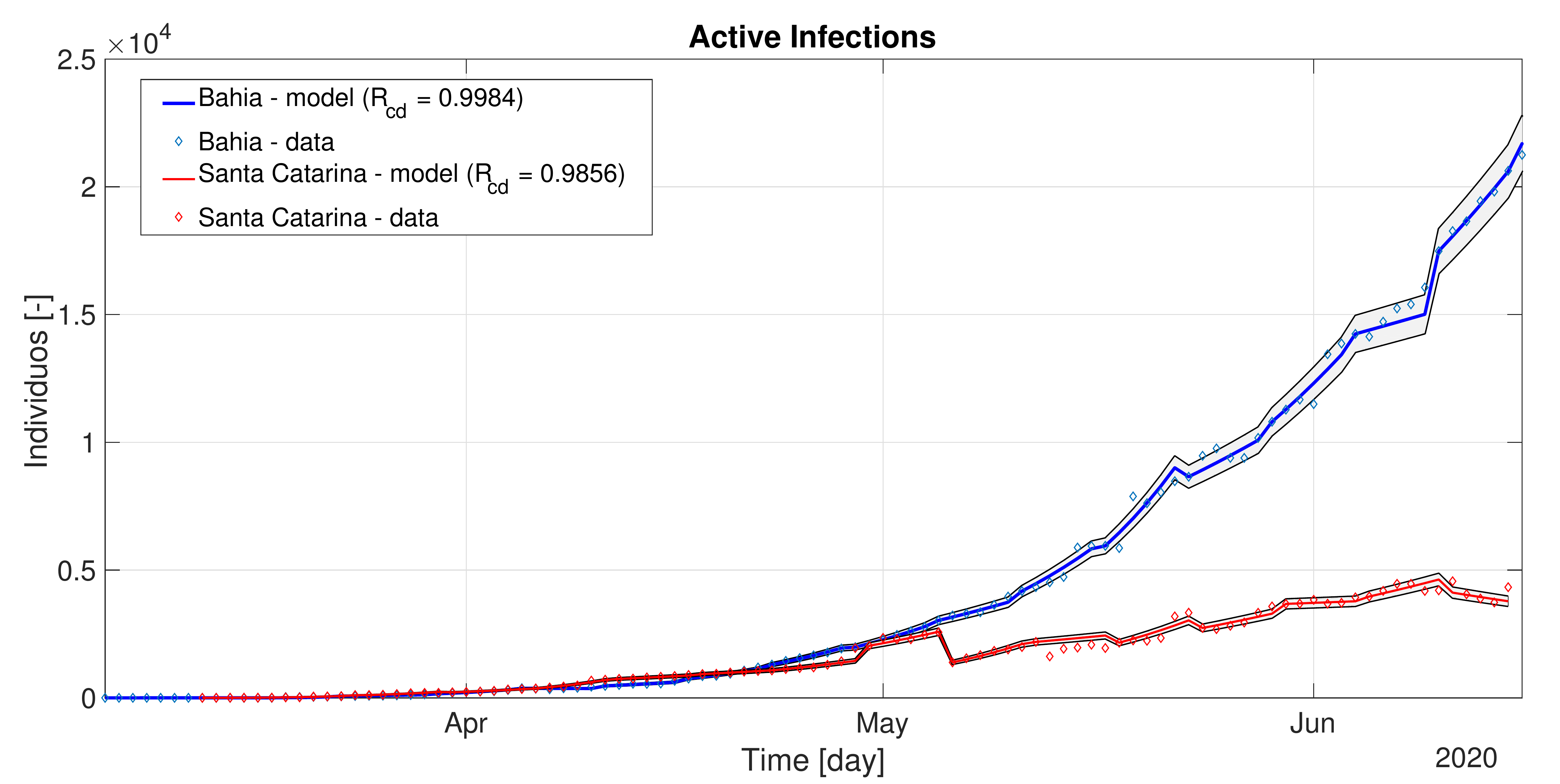}
                \caption{Identification Procedure: Active Infections ($95\%$ Confidence Interval)}
	\label{indetifI}
\end{figure}

\begin{figure}[htb]
	\centering
		\includegraphics[width=\linewidth]{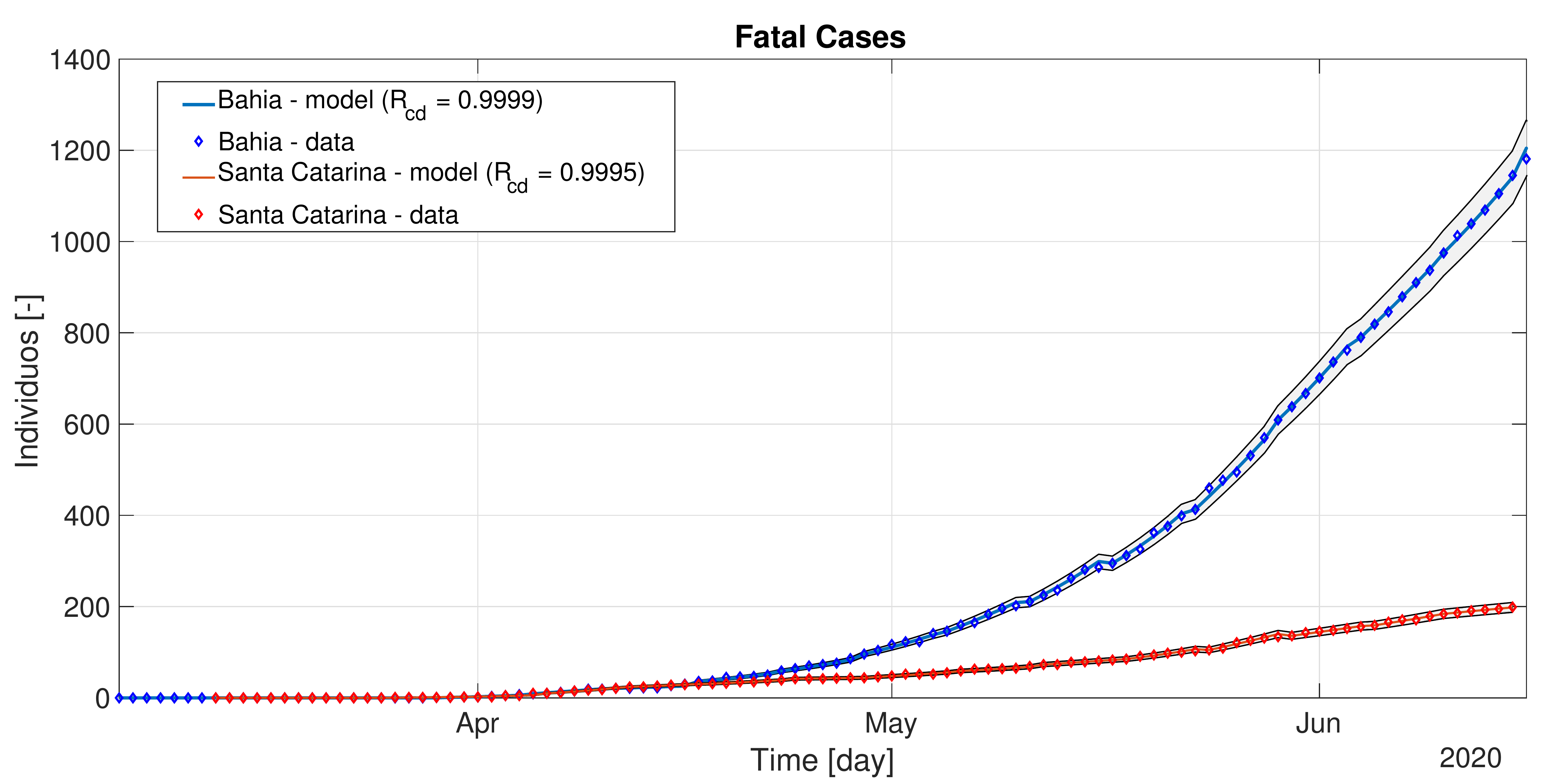}
                \caption{Identification Procedure: Fatal Cases  ($95\%$ Confidence Interval)}
	\label{indetifD}
\end{figure}

\begin{figure}[htb]
	\centering
		\includegraphics[width=\linewidth]{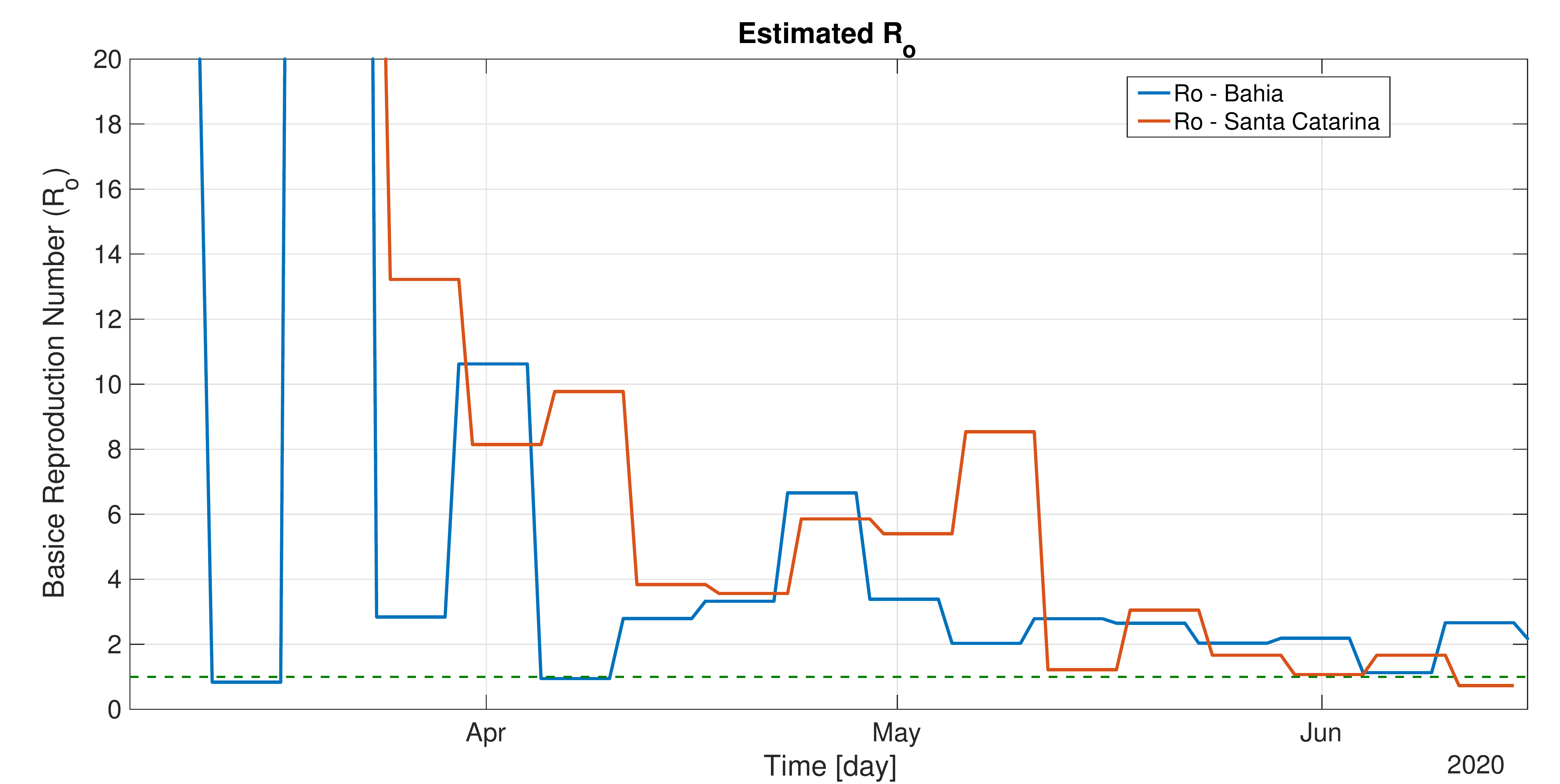}
                \caption{Identification Procedure: Smallest $R_0$, with respect to the previously shown $95\%$ Confidence Interval}
	\label{indetifR0}
\end{figure}

The obtained model parameters for the last data-set window, from June 10, 2020 to June, 16 2020 are presented in Table \ref{ModelParametersTable} (considering a full data window). Furthermore, Figures \ref{indetifI}, \ref{indetifD} and \ref{indetifR0} depict the model-data fitting results for the data-sets, regarding $I$, $D$ and $R_0$. Evidently, the identification procedure yields quite good parameter estimates, since the simulated SIRD model with parameter estimated for windows of $t_{opt} \,=\,$ 6 days globally represents the data very well, which can be observed from a coefficient of determination $R_{cd}$ very close to 1.

It is worth mentioning the different pandemic moment in each state. From Figures \ref{indetifI} and \ref{indetifR0}, it can be seen that SC has reduced the number of active infections, which is also reflected in $R_o <1$, representing that, at least at that moment, the disease reached its peak of active cases. On the other hand, the state of BA shows an increasing trend and also $R_o>1$, in which it can be expected that the number of cases will still grow.

\begin{table}[htbp]
    \caption{\label{ModelParametersTable}  Model Parameters.}
    \centering
\begin{tabular}{|c|c|ccc|}
\hline
\multicolumn{2}{|c|}{States} & \multicolumn{1}{c}{$\beta$} & \multicolumn{1}{c}{$\gamma$} & $\alpha$ \\ \hline
BA & \multirow{2}{*}{\begin{tabular}[c]{@{}c@{}}From June 10th\\  to June 16th\end{tabular}} & 0.181 & 0.053 & 0.017 \\ \cline{1-1} \cline{3-5} 
SC &  & 0.087 & 0.737 & 0.010 \\ \hline
\end{tabular}
\end{table}

 We note that the SIRD model parameters used for control are those for the last available window, as given in Table \ref{ModelParametersTable}. Since a window of $6$ days is shown to be sufficient to estimate the SIRD model parameters with model-fitting efficiency ($R_{cd}$ coefficient close to $1$), the most adequate control procedure is to  adjust the model of the MPC controller in an iterative fashion, as time progresses. This kind of procedure allows one to incorporate the variability of the SIRD parameters, which is inherent to the SARS-CoV-2 viral spread dynamics. We cannot proceed with this procedure since we consider the control action being deployed through the future (for which we have no data). Thus, we simply keep the last available SIRD parameters as those used for control.

\subsection{Control Results}

Considering the given parameters for the SIRD model, the control results are presented. The values for the social isolation response dynamics of Eq. \eqref{eq:Psi} are borrowed from \citep{morato2020optimal}. The maximal social isolation factors, as presented in Table \ref{tabelPsiEDO}, were retrieved from recent technical notes from these states.

\begin{table}[htb]
    \caption{\label{tabelPsiEDO} Social Isolation Response.}
    \centering
	\begin{tabular}{|c | c c |} 
		\hline 
        State & $\varrho$ & $\psi_{\infty}$ \\ \hline
        BA & $1.66$ days & $0.563$  \\ \hline
        SC & $1.66$ days & $0.514$  \\ \hline
    \end{tabular}
\end{table}

The MPC strategies are synthesized with a prediction horizon of $30$ days. This is coherent since the incubation period of the SARS-CoV-2 virus is of, at most, $14$ days. The weights $q_I$ and $q_u$ are chosen, respectively, as $0.5$ each, so that the MPC tries to find a ``balance" between minimizing infections and relaxing quarantine measures.

The achieved results are obtained considering an initial condition with respect to the available data at $11/06/2020$. We note, as of this date, BA has many infections (the ICU beds at BA are almost full), while SC has already passed through the infection peak. The control strategy is assumed to act from $12/06$ to mitigate the backlashes. The results indicate what \textbf{could still be done} to avoid catastrophic if no stronger health policy is employed. Through the sequel, OL denotes the results in "open-loop", i.e. without control $u = 0$ and, thus, with a social isolation factor $\psi \to 0$.

\begin{figure}[htb]
	\centering
		\includegraphics[width=\linewidth]{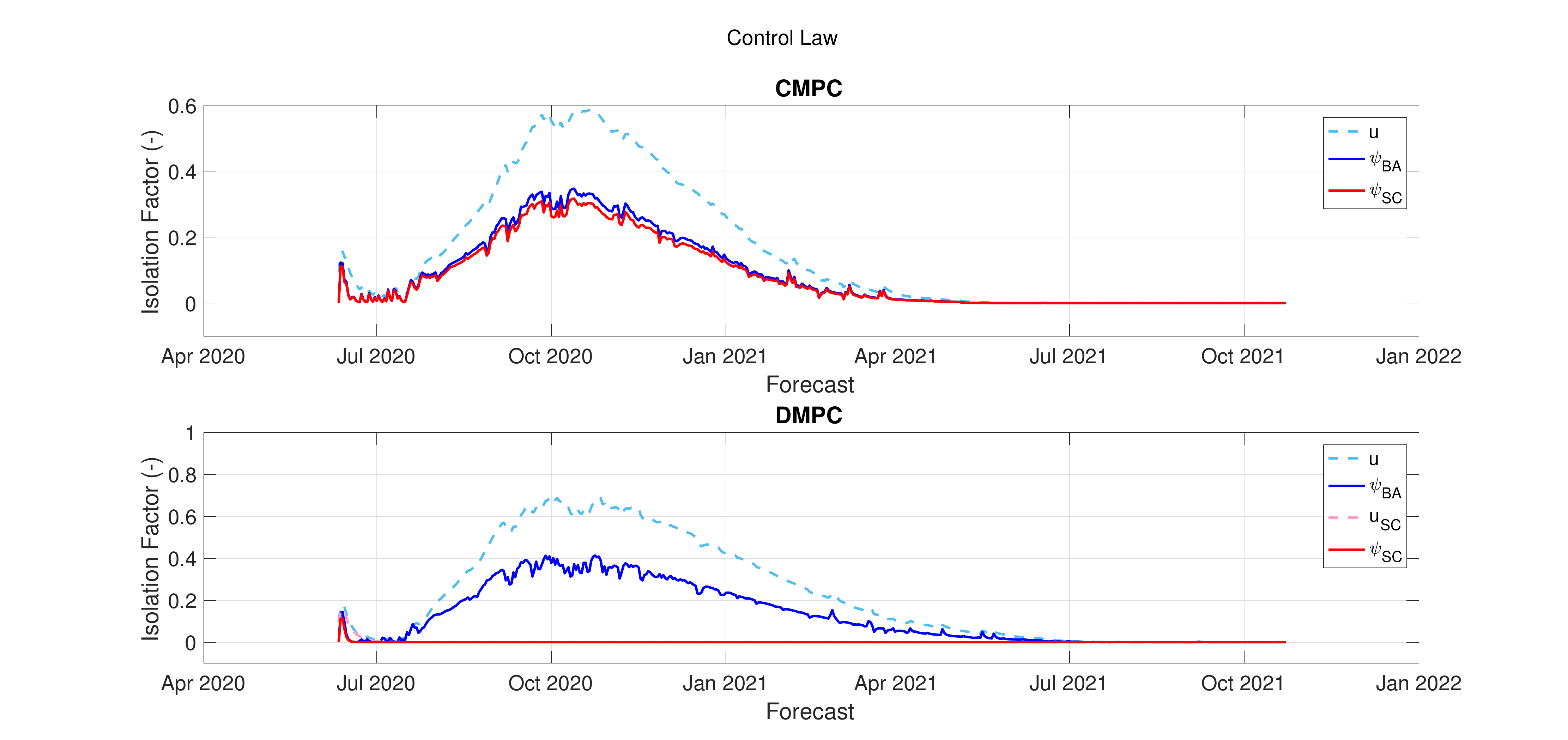}
                \caption{Control Results: Control Action and Social Response (MPCs are based on the nominal model)}
	\label{MPCu}
\end{figure}

In Figure \ref{MPCu}, the derived control laws are presented and compared to the social isolation factors $\psi$. The CMPC strategy is presented on the upper sub-plot, while the DMPC is shown below. Both strategies have similar behaviours with respect to BA, trying to suppress the spread of the virus by increasing the quarantine "strength" as the infections increase, and relaxing it afterwards; the forecast to the end of social isolation policies in BA is for June, 2021. Since SC shows an already decaying infection curve, the DMPC takes into account this specificity and indicates a relaxation much before, around August 2020. We note that the CMPC, since it considers both states, must determine a stronger policy to SC due to the elevated infection rate at BA, while the DMPC approach is able to individually plan the isolation, as expected. 

It must be stressed that we analyse the SIRD models as if there is no coupling effects between them. Anyhow, in practice, the $26$ states in Brazil cannot pursue individual social isolation laws (as the DMPC approach) since their borders are not closed. The DMPC results only indicate that local conditions should be taken into account, but a centralized coordination (like the CMPC) is forcefully necessary to reduce the infections all over Brazil. It seems to us much more prudent if the federal government dispatch a coordinated social distancing health policy (following a CMPC method), while each state figuring out their possible relaxations according to a DMPC approach and taken into account the infection level in the frontiers states. It does not seem reasonable to relax social isolation in SC by August $2020$ and expect that there is no migration/transit between people from neighbouring states (as Paraná or Rio Grande do Sul), which show much greater infection levels (and are previewed to relax quarantine much later).

\begin{figure}[htb]
	\centering
		\includegraphics[width=\linewidth]{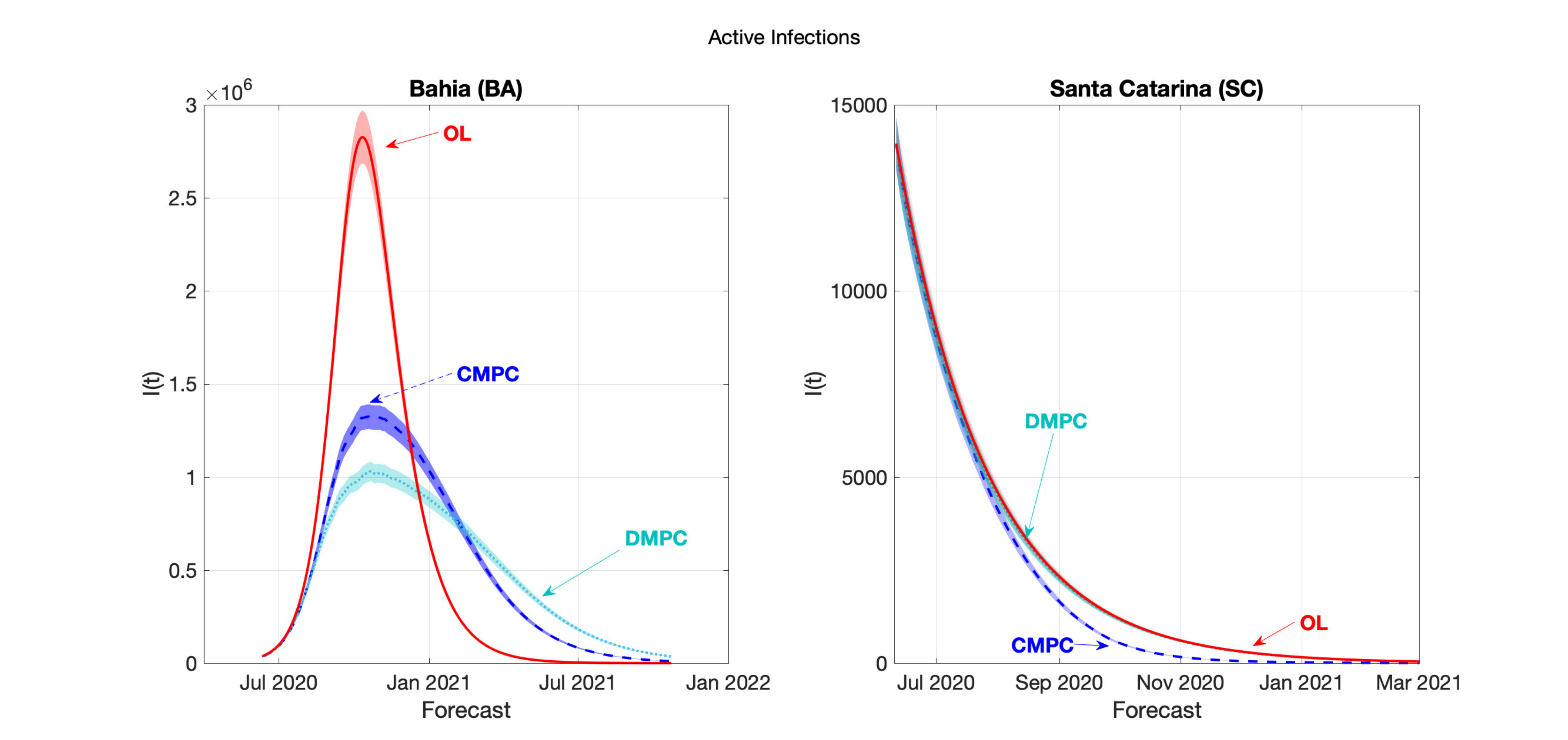}
                \caption{Control Results: Active Infection ($95\%$ Confidence Interval)}
	\label{MPCI}
\end{figure}

\begin{figure}[htb]
	\centering
		\includegraphics[width=\linewidth]{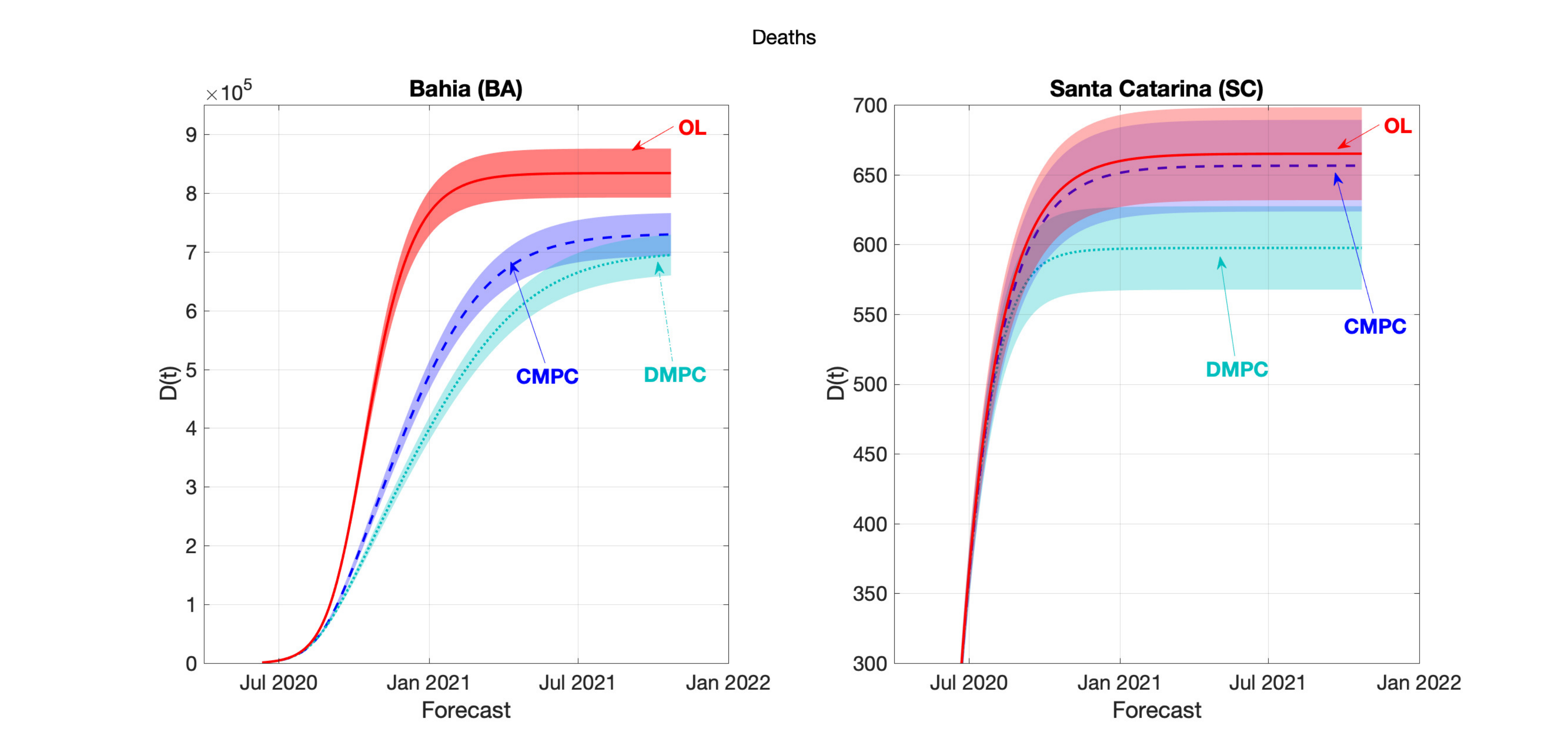}
                \caption{Control Results: Deceased Individuals  ($95\%$ Confidence Interval)}
	\label{MPCD}
\end{figure}

With respect to the depicted control laws, Figures \ref{MPCI} and \ref{MPCD} show, respectively, the evolution of the active infections and the total number of deaths due to COVID-19 in both states, over time. The results indicate that \textbf{over $100000$ lives could still be saved} in BA and \text{$70$} lives in SC, with respect to an OL condition. The amount of deaths in an OL scenario for BA are astounding. Of course, each life matters and this catastrophe is a lot to bare. Psychological and social traumas will mark the country. A hard isolation and a coordinated social distancing action could still be able to save many lives.

As a final (yet strikingly important) comment, we must discuss that this work only sketches preliminary results on how optimal control can be formalized for pandemic scenarios. An actual application of the proposed method (either CMPC or DMCP) depends on how the control signal can be translated into actual public health policies to be put into practice. This can be understood as some kind of actuator filtering of the control signal, since abrupt daily variations on $u$ make no sense regarding health policies. As an example, one cannot expect to determine relaxations (allow public transport) in one day to revert it in the following. The paradigm to consider only two states (total lockdown or total release) has been previously studied \citep{morato2020optimal} and also offers an elegant solution, but it seems that the preferable way to follow is to determine discretized values for $u$, which can be converted directly into practicable health policies. This kind of control signal is to be considered in future works, yet an easy route is to adequately filter the control signal generated with the proposed methods in this paper.

%\begin{figure}[htb]
%	\centering
		%\includegraphics[width=\linewidth]{./img/recovered.e%ps}
         %       \caption{Control Results: Recovered Individuals}
	%\label{MPCR}
%\end{figure}

\section{Discussions and Conclusions}
\label{sec6}
In this brief article, it is investigated how predictive control, optimization-based procedures could be used to formulate social isolation guidelines for the COVID-19 pandemic in Brazil, taking into account the spread of the virus in the states of Bahia and Santa Catarina. Centralized and distributed MPC approaches based on SIRD models with parameters identified via Least-Square optimization were proposed in this work. The results indicate that strong quarantine/lockdown measures still have to be enacted for some months before any relaxations can be though of.

Below, we summarize some key points:
\begin{itemize}
    \item The results corroborate the hypothesis formulated in \cite{hellewell2020feasibility} and previously discussed in \cite{morato2020optimal}, which indicate that herd immunity cannot be considered as plausible solution for Brazil, offering great risk and leading to elevated fatality due to multiple social-economical issues of the country.
    \item The control results show that a centralized, coordinated federal government action is necessary to set guidelines to the states, which can performed individual optimization procedures to determine when to relax quarantine measures. A forecast is presented which indicates that a coordinated social isolation public policy could save over $100000$ lives in just in these two states.
    \item The SARS-CoV-2 contagion is an inherently complex phenomenon and is influenced by many factors and exact prediction of the future dynamics is not possible and, therefore, the quantitative results presented herein cannot be account for without taking into account the uncertainty margins. Anyhow, the qualitative results are strong. The most correct control procedure should be based on a recurrent (daily) model tuning and re-calculation of the control law.  Since the country as been experiencing an unwillingness to formally start harder social isolation measures \citep{THELANCET20201461}, the social and economic costs of the pandemic might be brutal.
\end{itemize}

\section*{Acknowledgment}
The Authors acknowledge the financial support of National Council for Scientific and Technological Development (CNPq, Brazil) under grants $304032/2019-0$ and $201143/2019-4$ (PhD Program Abroad). M. M. M. and J. E. N. thank Saulo B. Bastos and Daniel O. Cajueiro for previous collaborations and discussions.

%\clearpage

%\section*{References}
\bibliographystyle{model5-names}
\bibliography{ifacconf}

\end{document}